\documentclass[useAMS,usenatbib]{mn2e}
\usepackage{graphicx}
\usepackage{lscape}
\usepackage{ulem}

\def\cm2{cm$^{-2}$}
\def\hkpc{$~h_{70}^{-1}$ kpc}
\def\icm{cm$^{-2}$}

\def\mgtwo{Mg~II}
\def\nhi{N(H~I)}

\title[Emission Line Abundances of Absorption Selected Galaxies at $z<0.5$]
{Emission Line Abundances of Absorption Selected Galaxies at $z<0.5$}
\author[Sara L. Ellison, Lisa J. Kewley and Gabriela Mall\'en-Ornelas]
{Sara L. Ellison$^{1}$\thanks{Email: sarae@uvic.ca}, 
Lisa J. Kewley$^{2}$\thanks{Current address: Institute 
for Astronomy, Honolulu, Hawaii, USA}
Gabriela Mall\'en-Ornelas$^{2}$, 
\\
$^1$University of Victoria,3800 Finnerty Rd, Victoria, BC, V8P 1A1, Canada\\
$^2$Harvard-Smithsonian Center for Astrophysics, 60 Garden Street, 
Cambridge, MA 02138 USA\\
}

\begin{document}

\maketitle

\begin{abstract}
We have obtained optical spectra of four galaxies associated with \mgtwo\
QSO absorbers at redshifts $0.10 < z < 0.45$.  We calculate the gas-phase 
oxygen abundance of these galaxies using the empirical R$_{23}$ strong 
line method.  The absolute $B$-band magnitudes of the galaxies span 
$-20.6 < {\rm M_B} < -18.3$.  
If the metallicities lie on the R$_{23}$ upper branch ($8.4 < 
{\rm log (O/H) + 12} <  8.9$), then the metallicities of
these absorption selected galaxies span the range
between 0.5--1.4 $Z_{\odot}$ and would 
be consistent with the well-known luminosity-metallicity relation 
for $0.10 < z < 0.45$ emission-line galaxies. However, such 
metallicities would be 0.5--1.0~dex higher than 
those observed in  damped Lyman $\alpha$ systems (DLAs) via
absorption line measurements at similar redshifts.  
Conversely, the lower R$_{23}$ branch calibration yields
metallicities $Z \sim 1/7 Z_{\odot}$, consistent with the DLA 
absorption metallicities at low redshifts.  In this case, 
the absorption selected galaxies would lie significantly lower 
than the luminosity-metallicity relation 
for emission-line galaxies at $z<0.5$.  
We discuss the implications and possible solutions for each scenario.

\end{abstract}


\section{Introduction}

Measuring the chemical content of galaxies at early cosmic epochs
is an important probe of the history of star formation and the
early evolutionary phases of galactic objects.  Although 
galaxy detections are now being pushed out to $z \sim 6$ 
and beyond (e.g., Hu et al.\ 2002; Dickinson et al.\ 2003a; Stanway,
Bunker \& McMahon
2003; Cuby et al.\ 2003), only a few high redshift objects
are bright enough to be studied in detail (e.g., Pettini et al.\ 2001, 2002; 
Teplitz et al.\ 2000; Kobulnicky \& Koo 2000; Steidel et al. 2004).  
Our direct
knowledge of the chemical evolution of galaxies is therefore mostly
limited to $z < 1$ where optical nebular emission lines can be used to
determine [O/H] (e.g., Kobulnicky \& Zaritsky
1999; Kobulnicky et al.\ 2003; Lilly, Carollo \& Stockton 2003; Maier,
Meisenheimer \& Hippelein 2004; Kobulnicky \& Kewley 2004).

Quasar absorption line studies provide an alternative means to measure
abundances out to arbitrarily high redshifts.  In particular, 
damped Lyman $\alpha$ (DLA) 
systems provide a powerful tracer of the chemical history of high redshift
galaxies (e.g., Pettini et al.\ 1997; Prochaska \& Wolfe 2002).  
DLA studies imply that there has been very little evolution in their weighted 
metallicities between $1.5 < z < 4$, with a mild increase by a factor 
of a few at lower redshifts 
(Pettini et al.\ 1999; Kulkarni \& Fall 2002; Prochaska et al.\ 2003;
Khare et al. 2004).  
This result is somewhat surprising given the rapid build-up in the stellar
mass density at $z<2$ (Dickinson et al.\ 2003b; Rudnick et al. 2003) 
and high star formation
rates at early times (e.g., Giavalisco et al.\ 2004 and references therein).

Observational bias has often been invoked in order to explain the 
apparent lack of observed metals.  For example, it has been suggested that 
dust may be responsible for obfuscating QSOs behind the most evolved galaxies
(Fall \& Pei 1993; Hou, Boissier \& Prantzos 2001; Churches et al.\ 2004).
In an attempt to circumvent this bias, Ellison et al.\ (2001) conducted
a DLA survey based on a radio selected sample of QSOs with complete optical
identifications. They found at most a factor of two difference in the mass
density of neutral gas and number density of DLAs at $z > 2$ compared
with optically selected QSO surveys. There is a similarly
small effect for \mgtwo\ absorption systems at lower redshifts, at least down
to $z \sim 0.7$ (Ellison et al., 2004a).  Given the
good agreement in number density and neutral gas content, Ellison (2000)
calculated that the metallicities of DLAs at $z \sim 2.5$ would have 
to be $> 0.5 Z_{\odot}$ to account for the mass density
of metals predicted from star formation rates (Pettini 1999).  
However, recent high resolution observations of the unbiased DLA sample
of Ellison et al. (2001) confirm that their metallicities
are in good agreement with the range of
extant abundance measurements, i.e. $1/10 - 1/15 Z_{\odot}$ (Akerman et al.,
in preparation).  The same is probably also true of 
lower redshift absorbers (Ellison et al., 2004a).  The results from
these 'complete' surveys have therefore not provided any 
support for the hypothesis that dust bias is responsible for the
lack of metallicity evolution in DLAs.

In this paper, we present the first results of a project whose aim
is to explore the possibility that absorption systems such as DLAs may not 
trace the bulk of the metals' reservoir at a given epoch.
We would like to address two questions:  
1)  what is the range of emission line metallicities in absorption 
selected galaxies
at a given redshift and 2) does the absorption line
abundance of a given absorption selected galaxy 
agree with its emission line metallicity?  This latter issue
has been the focus of considerable debate for the last decade
(e.g. Kunth et al. 1994; Kunth, Matteucci \& Marconi 1995; Pettini \& 
Lipman 1995),
and has seen a recent renewed interest in the literature
(Aloisi et al. 2003; Lebouteiller et al. 2004; Lecavelier des 
Etangs et al. 2004).

Here, we make a first step to addressing the former of these questions by 
measuring the HII region oxygen abundances in a small sample of
\textit{absorption selected} galaxies at $z<0.5$ and discuss
future observations that will tackle the latter issue. 

We assume cosmological parameters of $\Omega_M = 0.3$, 
$\Omega_{\Lambda}=0.7$, $H_0 = 70\,h_{70}$\,km~s$^{-1}$~Mpc$^{-1}$.  In this
cosmology, 1 arcsec transverse on the sky corresponds to 1.8 \hkpc\
at $z=0.1$ and 5.4 \hkpc\ at $z=0.4$.

\section{Sample Selection, Observations \& Data Reduction}


\begin{table*}
\caption{Observations and Galaxy/Absorber Parameters} 
\begin{tabular}{lcccccccc}
\hline
Field & Galaxy & QSO--Galaxy & Galaxy & \nhi & Grism & Slit width & Exposure &
Resolution \\
 & Redshift & Sepn (arcsec) & $M_B^{a}$ & atoms \icm & & (arcsec) & (seconds) & FWHM (\AA) \\
\hline
0150$-$203 & 0.383 & 10.5  & $-19.52$ & 3$\times10^{18 c}$& Red & 1.0 & 8700 & 13 \\
0151+045A$^{b}$  & 0.160 & 6.4   & $-19.43$ & 7$\times10^{19 c}$& Blue &1.4 & 3600  & 17 \\
0151+045B$^b$  & 0.160 & 10.9  & $-18.32$ &  7$\times10^{19 c}$& Blue &1.2 & 6300 & 14  \\
0229+131   & 0.417 & 6.8   & $-20.58$ &...& Red &1.0 & 4200  & 13 \\
\hline
\end{tabular}
\\
a: Calculated from the $m_r$ values in Guillemin \& Bergeron 
(1997) using $k$-correction templates from Sawicki, Lin \& Yee (1997) 
in our cosmology (assuming an Sbc type).
b:  Two galaxies with redshifts consistent with that of 
the absorber are found in this field with small impact parameters.  It is 
not clear which is causing the absorption in the QSO.
c: S. Rao, private communication, 2004.
\end{table*}


\begin{figure*}
\centerline{\rotatebox{270}{\resizebox{15cm}{!}
{\includegraphics{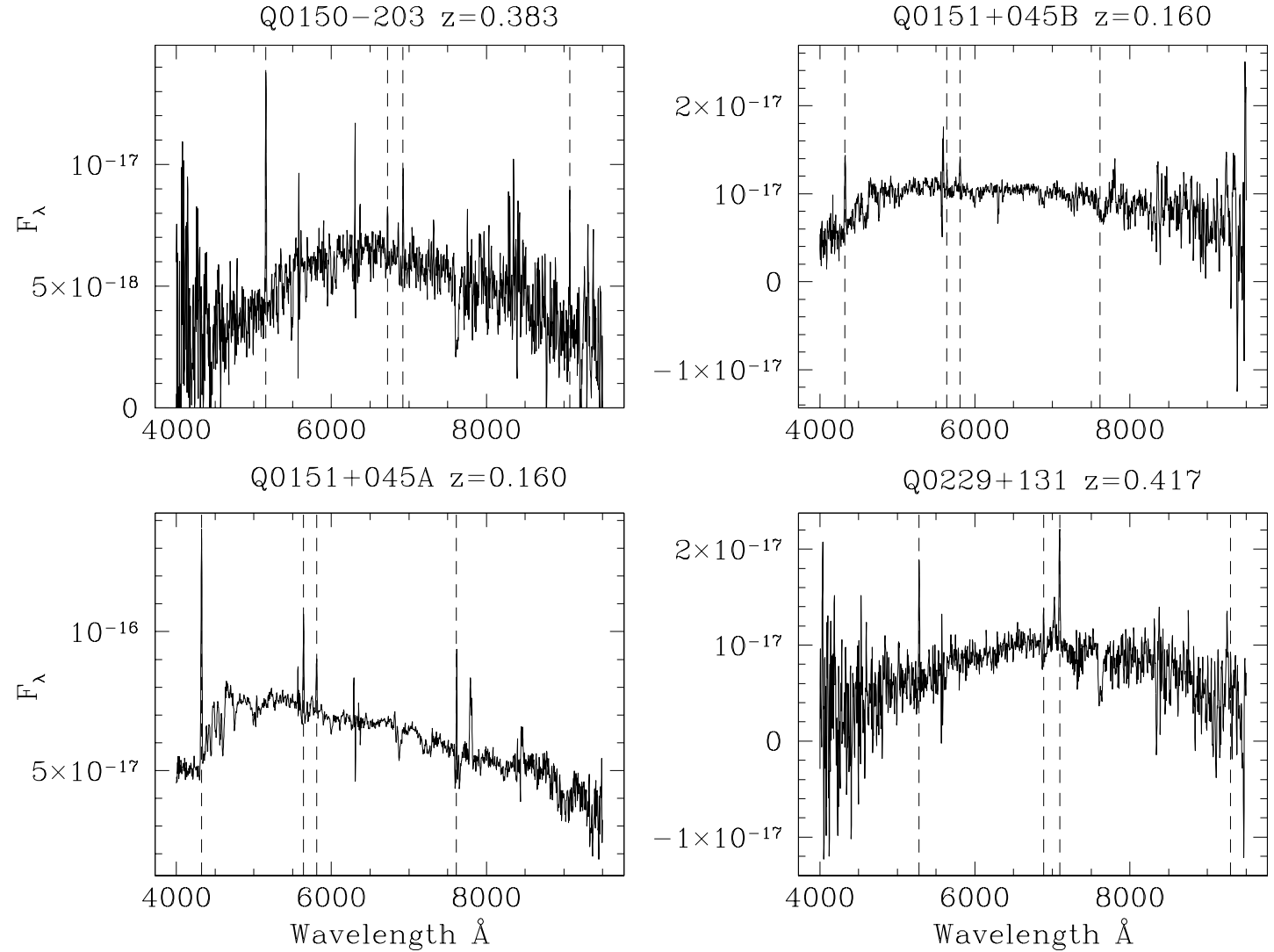}}}}
\caption{Galaxy spectra with emission lines marked from left to right:
[O~II] $\lambda$3727, H$\beta$, [O~III] $\lambda$5007 and H$\alpha$. }
\end{figure*}

Since very few DLA galaxies have been spectroscopically confirmed,
we chose to begin our study with a sample of galaxies associated
with strong \mgtwo\ absorbers selected from the compilations of
Guillemin \& Bergeron (1997) and Bergeron \& Boiss\'e (1991). 
An advantage of \mgtwo\ selection is that these absorbers
are more easily identified in QSO spectra at low redshift. The
\mgtwo\ doublet occurs at a rest wavelength of $\lambda\lambda$
2796, 2803 \AA\ and is therefore observable in the optical regime
for $0.3 < z < 2.5$.  DLAs, on the other hand,
require UV spectroscopy to confirm their \nhi\ at $z<1.6$
and are consequently less well studied.

Table~1 lists the fields for which we performed galaxy spectroscopy;
precise galaxy coordinates can be found in Guillemin \& Bergeron
(1997) and Bergeron \& Boiss\'e (1991). One of the \mgtwo\
systems in our sample, towards Q0151+045, has two candidate
absorbing galaxies; we have obtained spectra of both galaxies.
Our sample therefore consists of three absorption systems with
spectroscopy for four galaxies.
The H~I column densities for 2/3 of the absorbers 
can be measured from existing HST/FOS
archival spectra (Rao \& Turnshek 2000; S. Rao, 2004,
private communication).  
For the third QSO, Q0229+13, the presence of a higher
redshift Lyman limit system precludes the measurement of \nhi\
at $z=0.417$.  We note that the selection of bright galaxies 
with confirmed spectroscopic redshifts could potentially bias 
this sample towards higher
metallicities.  Given that \mgtwo/DLA galaxies can exhibit a
range of luminosities and morphologies (Steidel, Dickinson \& Persson 
1994; Chen \&
Lanzetta 2003; Rao et al.\ 2003), a more extensive survey that
includes lower-luminosity systems may reveal a wider range of galaxy
metallicities than exhibited by our current sample.

Spectra were obtained using LDSS2 at the Magellan II telescope
during runs scheduled between 2003 September 28--30 and November 8--11 
under conditions of mixed transparency.  
Observations were executed at or near the parallactic angle in order
to avoid losses which could otherwise introduce significant systematic
errors into our flux ratios.     We used the medium red and
blue grisms with slit widths of between 1.0 and 1.4 arcsecs,
yielding FWHM resolutions of between 13 and 17 \AA, as detailed in Table 1.
In addition to the galaxies listed in Table 1,
we observed the object which Guillemin \& Bergeron (1997) identify as
the galaxy causing
absorption towards Q0302$-$223.  However, our spectrum reveals
this object to be stellar and we consider it no further here.

Data were reduced using standard IRAF\footnote{IRAF is distributed by
the National Optical Astronomy Observatories, which is operated by the
Association of Universities for Research in Astronomy, Inc. (AURA)
under cooperative agreement with the National Science Foundation}
packages; the 2D spectra were bias subtracted, flat-fielded, cleaned
of cosmic rays using L.A. Cosmic (van Dokkum 2001) and finally
calibrated for wavelength and flux.  The final spectra are presented
in Figure 1.  In order to measure the emission line fluxes, we
extracted 1D spectra, produced an average for each galaxy and fit
single or double Gaussians using IRAF's $splot$ (metal lines) and
$ngaussfit$ (Balmer lines) taking into account the underlying
stellar absorption.  IRAF's $ngaussfit$ allows simultaneous
fitting of blended absorption and emission components.  First,
the continuum around the line is defined and then the emission and
absorption are simultaneously fitted through an iterative process by
varying the 5 parameters which define each component (continuum
zero point, slope, Gaussian amplitude, central wavelength and FWHM).
The velocity widths of lines of the same element were tied together
in this fitting process.   We measured the emission line fluxes of
[O~II] $\lambda$3727, H$\beta$ and [O~III] $\lambda$5007
for each galaxy. The observed fluxes with Balmer corrections are 
given in Table 2.  Although we detect H$\alpha$ 
in most cases, accurate flux measurements of both H$\alpha$ and
[N~II]$\lambda$6584 were generally not possible due to either 
nearby telluric absorption or low S/N.  


\begin{table*}
\caption{Observed Galaxy Line Fluxes}
\begin{tabular}{lcccc}
\hline
Field & \multicolumn{4}{c}{Line Fluxes$^a$}  \\  
[.2ex] \cline{2-5}
& [O~II]$\lambda$ 3727 & H$\beta$ & [O~III]$\lambda$ 5007 & H$\alpha$  \\
\hline
0150$-$203 & 1.24$\times 10^{-16}$ & 4.92$\times 10^{-17}$ & 4.64$\times 10^{-17}$ &  6.43$\times 10^{-17}$  \\
0151+045A  & 1.46$\times 10^{-15}$ & 6.47$\times 10^{-16}$ &  2.65$\times 10^{-16}$ & ... \\
0151+045B  &  1.83$\times 10^{-16}$ &  3.89$\times 10^{-17}$ &  7.51$\times 10^{-17}$ & ... \\
0229+131   & 1.55$\times 10^{-16}$ &  9.45$\times 10^{-17}$ &  1.95$\times 10^{-16}$ & ... \\
\hline
\end{tabular}
\\
a:  Values are for relative \textit{observed} line fluxes in units 
of ergs s$^{-}$ cm$^{-2}$ \AA$^{-1}$, except for Balmer lines which
have been corrected for underlying stellar absorption.
\end{table*}


\begin{table*}
\caption{Extinction Corrected Galaxy Line Fluxes}
\begin{tabular}{lccccc}
\hline
Field & \multicolumn{4}{c}{Line Fluxes$^a$} & Log R$_{23}$ \\  
[.2ex] \cline{2-5}
& [O~II]$\lambda$ 3727 & H$\beta$ & [O~III]$\lambda$ 5007 & H$\alpha$ & \\
\hline
0150$-$203 & 2.62$\times 10^{-16}$ & 8.66$\times 10^{-17}$ & 7.83$\times 10^{-17}$ &  9.29$\times 10^{-17}$ & 0.63 \\
0151+045A  & 3.08$\times 10^{-15}$ & 1.14$\times 10^{-15}$ &  4.47$\times 10^{-16}$ & ...& 0.51 \\
0151+045B  &  2.97$\times 10^{-16}$ &  5.61$\times 10^{-17}$ &  1.05$\times 10^{-16}$ & ... & 0.89\\
0229+131   & 4.16$\times 10^{-16}$ &  2.03$\times 10^{-17}$ &  3.96$\times 10^{-16}$ & ...& 0.61 \\
\hline
\end{tabular}
\\
a:  Values are for relative extinction- and Balmer-corrected line fluxes 
in units of ergs s$^{-}$ cm$^{-2}$ \AA$^{-1}$.
\end{table*}

\section{Abundance Determinations}


\begin{figure}
\centerline{\rotatebox{270}{\resizebox{7cm}{!}
{\includegraphics{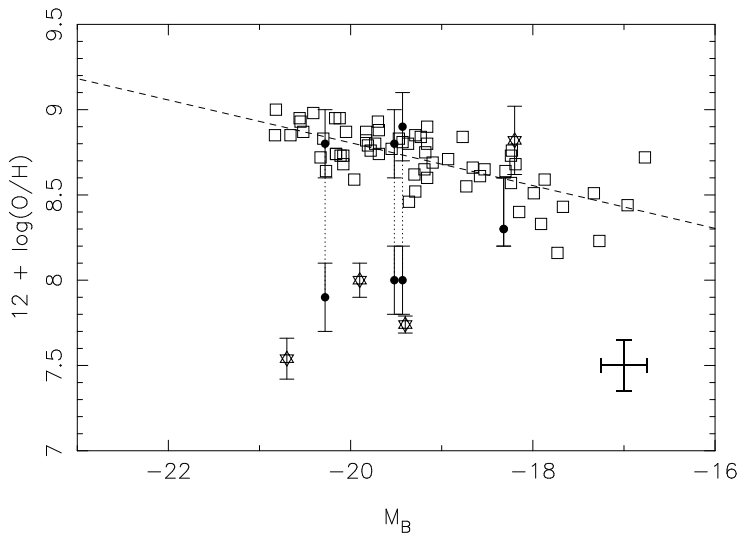}}}}
\caption{Luminosity-metallicity relation for $z<0.5$ galaxies taken from
Kobulnicky \& Kewley (2004; open squares, typical error bar shown
in bottom right corner) and this work (filled circles).
The dotted lines join the upper and lower $R_{23}$ branch metallicities 
(the range of possible turnover region metallicities is given by
the error bars in the case of 0151+045B) for a given galaxy in our sample. 
The dashed line is the least squares fit to the KZ99 data and has
the form 12 + log(O/H) = 6.30 $-$ 0.125 $M_B$.  
The stars represent the values listed in Table 5 for literature
DLAs with measured absorption abundances and galaxy counterparts.}
\end{figure}


\begin{table*}
\caption{Galaxy Abundances}
\begin{tabular}{lcccccccc}
\hline
Field & \multicolumn{3}{c}{log(O/H)+12, upper} & & \multicolumn{2}{c}{log(O/H)+12, lower} & log(O/H)+12 & log(O/H)+12 \\  
[.2ex] \cline{2-4} \cline{6-7}
 & M91 & ZKH94 & KD02 & & M91 & KD02 & upper adopted & lower adopted \\
\hline
0150$-$203 & 8.69 & 8.88 & 8.91 & & 7.92 & 8.11 & 8.8$\pm$0.2 & 8.0$\pm$0.2\\
0151+045A  & 8.75 & 9.00 & 9.00 & & 7.87 & 8.04 & 8.9$\pm$0.2 & 8.0$\pm$0.2\\
0151+045B$^a$  & 8.35 & 8.45 & 8.54 & & ... & ... & 8.4$^{+0.3}_{-0.1}$ & ...\\
0229+131   & 8.74 & 8.82 & 8.88 & & 7.81  & 8.05 & 8.8$\pm$0.2 & 7.9$\pm$0.2\\
\hline
\end{tabular}
\\
a:  Only a turnover region metallicity is quoted with 
error bars indicatative of the large uncertainty in 
metallicity. 
\end{table*}

We determine abundances using the R$_{23}$ method originally
formulated by Pagel et al.\ (1979) and recently re-calibrated by
Kewley \& Dopita (2002, KD02).  
We correct the observed emission line fluxes for dust extinction
by estimating a value of $E(B-V)$ and assuming the shape of the
extinction curve described by Cardelli, Clayton \& Mathis (1989) with
an adopted R$_V$=3.1.  In the absence of accurate $H \alpha$ fluxes, 
we use the local $E(B-V)$ to M$_B$
relation (Jansen et al.\ 2001) converted to our cosmology (R. Jansen,
private communication, 2003) to determine the extinction corrections for
0151+045A,B and 0229+131:

\begin{equation}
E(B-V) = -(0.0507\pm0.0068)\times M_B  - (0.818\pm0.127) 
\end{equation}

For 0150$-$203 we can measure the H$\alpha$/H$\beta$
ratio and, by assuming a theoretical value of 2.85 (for case B recombination
at T$=10^4$K and $n_{e} \sim
10^2 - 10^4\,{\rm cm}^{-3}$; Osterbrock 1989), 
directly derive $E(B-V)=0.17$ from the Balmer decrement.  This `direct'
determination of the extinction is precisely  
the value which is predicted by equation 1. Our final extinction
corrected (relative) line fluxes are given in Table 3.

To check for the presence of AGN we used the ratios of [O~III]$\lambda
5007$/H$\beta$ and [O~II]$\lambda 3727$/H$\beta$ as parametrized
by Lamareille et al. (2004).  All of our galaxies have line ratios 
that place them well within the starburst galaxy locus.

We calculate the value of $R_{23}$ (Pagel et al.\ 1979) after
extinction correction of fluxes.  The results are given in the
final column of Table 3, assuming that the flux
of [O~III] $\lambda$ 4959 = 1/3 $\times$ [O~III] $\lambda$ 5007.
We estimate the metallicity, log(O/H)+12, using three R$_{23}$
calibrations: McGaugh (1991, M91), Zaritsky, Kennicutt \& Huchra
(1994, ZKH94) and KD02.  The typical metallicity error associated with
any given calibration is typically $\sim$ 0.1 dex (see individual
references for details).  We note in passing that KD02 is the
only one of the three calibrations discussed here which solves
iteratively for the ionization parameter as well as oxygen abundance.  
All $R_{23}$ calibrations are double 
valued with respect to metallicity, resulting in two possible solutions,
the so-called `upper' and `lower' branches.
In Table 4 we list the metallicities for M91, ZKH94 and KD02
for both the upper and lower $R_{23}$ branches, except for
ZKH94 which was only calibrated for the upper branch.  The difference
between various published calibrations illustrates the typical
uncertainties in the empirical methods.  By trials with different
plausible Balmer absorption and extinction corrections we estimate that
these effects contribute a further 0.15 dex uncertainty.  We therefore 
allocate a final error of 0.2 dex to our adopted metallicities
(last two columns of Table 4) which are themselves averages of M91 and
KD02 (Kobulnicky \& Kewley 2004).  The adopted
metallicity error for 0150$-$203 is larger and encompasses the
local maximum of the $R_{23}$ calibration turnover (see below).

Our current spectra are insufficient to constrain the metallicities to 
either the upper or lower $R_{23}$ branch.  The [N~II]$\lambda$6584/H$\alpha$ 
ratio is commonly used to break the $R_{23}$ 
degeneracy (e.g. Maier, Meisenheimer \& Hippelein 2004).  Unfortunately, 
we only have a reliable H$\alpha$ flux for one galaxy and no 
trustworthy [N~II]$\lambda$6584 measurements.   One galaxy, (0151+045B) 
lies in the turn-around region of $R_{23}$;  
its metallicity is $8.3 < \rm{log(O/H)} + 12 \le 8.7$
(but note that this is one of two candidate galaxies for the absorber). 
For the remaining three galaxies, the metallicities may lie on the 
upper or lower $R_{23}$ branch. We discuss each possibility below.

\subsection{Metallicities on the upper $R_{23}$ branch}

If one or more metallicities of our Mg~II systems lie on the upper branch, 
those metallicities and luminosities would be consistent with the well-known 
luminosity-metallicity (LZ) relation for emission-line 
galaxies at similar redshifts (Figure 2).  We note that for one of the Mg~II
galaxies presented here (0151+045A), independent emission-line
spectroscopy
which includes a nitrogen detection indicates that, at least for
this one galaxy, the upper $R_{23}$ calibration is appropriate
(Christensen et al. 2004).  In Figure 2 we show the upper and 
lower-branch metallicities of our 
Mg~II galaxies compared with the $z < 0.5$ emission-line galaxy LZ  
relation from Kobulnicky \& Kewley (2004).  

If our Mg~II system metallicities lie on the upper branch, then their 
metallicities would be significantly higher than the DLA absorption 
metallicities by 0.5 -- 1.0~dex.  Such a large discrepancy can not 
be explained by the differences between different emission-line 
diagnostics (0.1--0.3~dex; Garnett, Kennicutt \& Bresolin 2004; 
Kobulnicky \& Kewley 2004).    A  combination
of strong abundance gradients and considerable QSO--galaxy 
impact parameters is one potential 
solution to this dilemma.  Our spectra only cover the nuclear 
regions of the Mg~II systems, whereas the impact parameters 
correspond to $\sim$ 15 -- 50 \hkpc.  If a Mg~II system has a 
strong metallicity gradient, then emission-line and absorption-line 
abundances at these impact parameters could 
deviate significantly.  To further illustrate this
point, we have plotted LZ data in Figure 2 for four DLAs or 
sub-DLAs from the
compilation of Chen \& Lanzetta (2003).  For these
four absorbers we plot the metallicity based on DLA absorption
line measurements and the absolute $B$-band magnitude of
the identified galaxy counterpart (see Table 5 for galaxy and DLA
parameters).  We have converted the 
AB magnitudes given in Chen \& Lanzetta (2003) to the Vega scale 
and corrected for our adopted value of $H_0$.    
We follow Prochaska et al. (2003) in preferentially
using the abundance, [X/H]\footnote{In the usual notation, [X/H]=
log N(X) $-$ log N(H) $-$ log (X/H)$_{\odot}$.} 
[Zn/H] or [Si/H] if possible, otherwise adopting [Fe/H]+0.4
as an indication of metallicity.  As noted previously by Pettini et al.
(2000), Q0058+016, whose impact parameter is only 8 \hkpc, 
lies within the LZ range of field galaxies at
moderate redshifts.  The other three DLAs are significantly more
metal poor for their luminosity than field galaxies at $z<0.5$,
but their impact parameters are also substantial ($\sim 20$ \hkpc).
We are currently investigating quantitatively the effect
of abundance gradients with a combination of 
theoretical Monte Carlo simulations and detailed metallicity gradient 
data.  The results of this theoretical work will be published in a 
future paper.
 
To summarise this subsection, if $Z \sim Z_{\odot}$ nebular 
abundances are confirmed in
a larger sample of DLA galaxies with sub-solar absorption abundances,
this would be evidence that although the DLAs are representative of
gas cross-section, they may not trace the major repositories 
of metals, possibly due to substantial impact parameters and metallicity
gradients.
In one case (0151+045A) spectroscopy by Christensen et al. (2004)
agree with our upper branch metallicity and support an approximately
solar metallicity for this galaxy.

\subsection{Metallicities on the lower $R_{23}$ branch}

If one or more metallicities of our systems lie on the lower branch, 
those metallicities would be 
consistent with the DLA absorption metallicities at similar redshifts.
However, in this case, our Mg~II absorption selected galaxies 
would deviate significantly
(and uniquely amongst star-forming galaxies at this epoch)
from the star-forming LZ relation.  If the LZ relation is representative of 
the galaxy population 
as a whole at redshifts $0.1 < z < 0.5$, then it is unlikely that random 
intervening galaxies
in QSO sightlines would preferentially isolate a population of
galaxies with a unique LZ relation.   On the other hand, if the 
emission-line metallicities of our 
Mg~II systems lie on the lower $R_{23}$ branch, this would support 
the hypothesis that DLA 
absorption metallicities are representative of the galaxy population 
as a whole at redshifts 
$0.1 < z < 0.5$.   An implication of such a result would be that the 
emission-line LZ relation is not representative of the galaxy population.  
Such a scenario might hold if, for example, sample 
selection caused the LZ relation to be biased towards high metallicity 
galaxies.  


\begin{figure*}
\centerline{\rotatebox{270}{\resizebox{12cm}{!}
{\includegraphics{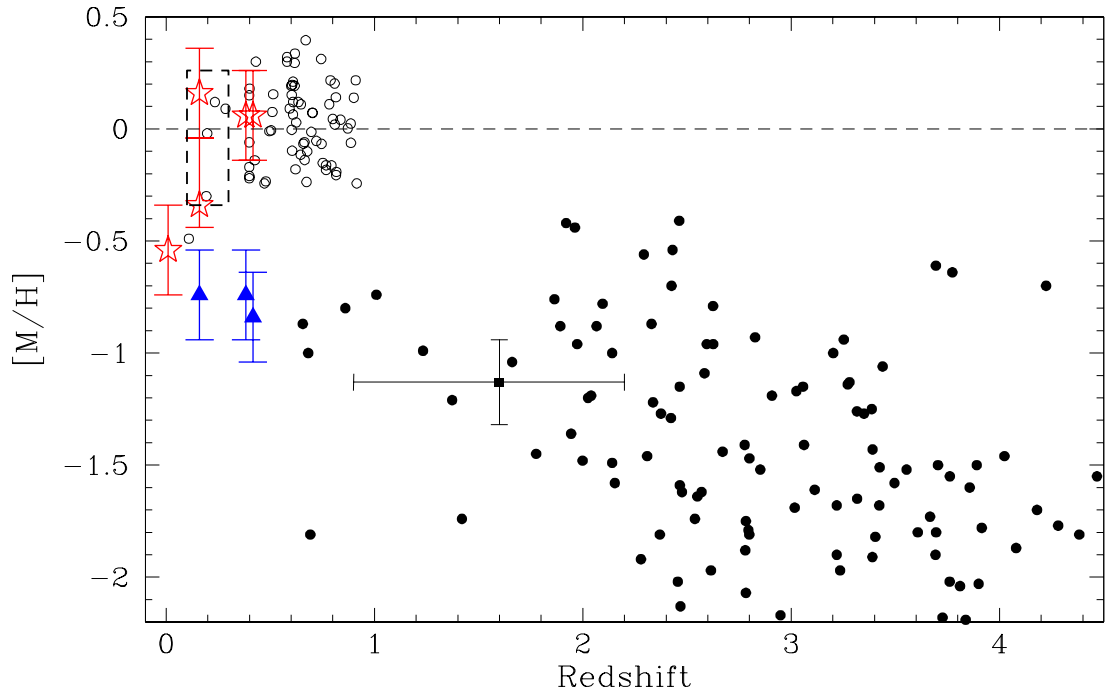}}}}
\caption{Metallicities of DLAs (solid circles) based on UV absorption 
lines in QSOs taken from Prochaska et al.\ (2003), emission line galaxies 
(open circles) taken from Kobulnicky \& Zaritsky (1999) and Lilly,
Carollo \& Stockton (2003) and absorption selected galaxies (open
stars/filled triangles)
from this sample (assuming upper/lower branch metallicities)
and Schulte-Ladbeck et al.\ (2004) (one 
DLA at $z \sim 0$).  Note that 0151+045B is in the turnover region
so its open star and error bars represents the range of possible
metallicies for this galaxy.
The dashed region shows the locus of the majority
of star forming SDSS galaxies with $0.1 < z < 0.3$ (Schulte-Ladbeck et
al. 2003) and the solid square is the metallicity of low redshift
DLAs based on the SDSS composite from Nestor et al. (2003).}
\end{figure*}


\begin{table*}
\caption{Properties of Literature DLA Absorbers and Galaxies} 
\begin{tabular}{lcccccc}
\hline

QSO & Galaxy & DLA & DLA & X & Metallicity & Impact \\
 & $M_B^a$ & redshift & [X/H]$^b$ & & Reference & Parameter\\
 & & & & & & (\hkpc) \\ 
\hline
Q0058+019   & $-18.2$ & 0.613 & $+0.08\pm0.21$ & Zn & Pettini et al. (2000) & 8.1 \\
Q0302$-$223 & $-19.9$ & 1.009 & $-0.73\pm0.12$ & Si & Pettini et al. (2000) & 26.6 \\
Q1122$-$165 & $-19.4$ & 0.682 & $-1.00\pm0.05$ & Fe+0.4 & Ledoux et al. (2002) & 25.2\\
Q1328+307 & $-20.7$ & 0.692 & $-1.20\pm0.12$ & Zn &  Ledoux et al. (2002)&  17.7\\

\hline
\end{tabular}
\\
a: Magnitudes taken from Chen \& Lanzetta (2003) and converted to our cosmology
b: Absorption line metallicity based on element X, with correction
of 0.4 dex in the case of Fe to account for dust depletion (e.g.
Prochaska et al. 2003).
\end{table*}

\subsection{Discussion}

Quasar absorption lines potentially offer an unbiased and
representative census of gas-phase metal abundance.  However, 
some observations indicate that random
sightlines through DLAs do not sample the regions where star
formation and chemical enrichment have been rife.   For example, there is
a serious shortfall between measured absorption abundances and the
predicted abundances from both star formation rates (Pagel 1999;
Pettini 1999 and recently re-assessed by Wolfe, Gawiser \& Prochaska 2003) 
and from smoothed particle hydrodynamic (SPH) galaxy
simulations (e.g., Nagamine, Springel \& Hernquist 2003). One
solution to this `missing metals' problem
is if a significant fraction of metals actually exist outside
the cool, diffuse gas phase. For example, the hot ISM
component may contain a significant fraction of a galaxy's metals.  
The local starburst
galaxy NGC 1569 has a metallicity of $Z=1/3 Z_{\odot}$ from HII
regions, but approximately solar abundance in a hot metal-loaded wind
(Kobulnicky \& Skillman 1997; Martin, Kobulnicky \& Heckman 2002).
Alternatively, Dunne, Eales \& Edmunds (2003) have claimed that most of 
the metals 
in high redshift galaxies are actually locked in the dust phase.
A more persistent and troubling concern is that
DLAs are observed to be metal-poor at all redshifts and show
little evolution with time (e.g. Pettini et al. 1999).
Therefore, regardless of the metals budget, DLAs consistently
represent regions which have experienced little chemical enrichment.
Given that dust bias can apparently be ruled out as the culprit
for this (Ellison et al. 2001; Ellison et al. 2004a; Akerman et al. 
in preparation) we are left to ponder
if \textit{absorption-selected galaxies are fundamentally metal-poor, or
whether random sight-lines preferentially probe metal poor gas}.

One way in which to address this question is to examine the
range of emission line metallicities at a given epoch and compare
with those determined from absorption lines.  In this work, we
have made a first step towards this goal by
estimating oxygen abundances 
for a small sample of galaxies selected on the basis of strong \mgtwo\ 
absorption in nearby (i.e., with impact parameters of 5--10 arcseconds) 
QSOs.  Figure 2 indicates that there is a conflict between the emission-line 
LZ relation and DLA metallicities in three out of four systems
(open squares, compared with open stars) between $0.1 < z < 0.5$.  
Clearly, there is no way to resolve the emission-line LZ relation with 
the DLA metallicities, regardless of whether our galaxies lie on the 
upper or lower branch.  The solution of the R$_{23}$ degeneracy could
potentially make a significant impact on either LZ relation studies, 
or on QSO absorption studies. 

In Figure 3, we show metallicity vs.\ redshift for our four
absorption-selected galaxies (two of which are candidates for the same
\mgtwo\ system, but this does not affect our conclusions), for the 
low-redshift DLA galaxy
SBS~1543+593 (Schulte-Ladbeck et al.\ 2004), for star-forming galaxies
at $z<1$ (KZ99; Lilly, Carollo \& Stockton 2003), and for a
sample of DLAs (Prochaska
et al.\ 2003).   We have included both upper and lower $R_{23}$
branch values, using the 'adopted' value from Table 4.  
If the upper branch calibration is appropriate, as tentatively
indicated from the LZ relation, our small sample of galaxies have
abundances $\sim$ 1 dex larger than low redshift DLAs.  If the
lower $R_{23}$ branch is appropriate, the galaxies have metallicities
consistent with the DLAs.  Figure 3 also emphasizes the
need for more low-redshift DLA abundance measurements.  We note that
the galaxies selected for this pilot sample are relatively bright,
so may represent the more metal-rich end of the distribution.

The galaxies in our sample are selected
based on \mgtwo\ absorption and at least 3 of them 
have H~I column densities
below the canonical limit for DLAs ($2\times10^{20}$ cm$^{-2}$).
However, P\'eroux et al. (2003) find that `sub-DLAs' (which
include absorbers with \nhi\ down to $10^{19}$ \icm)
have metallicities consistent with the higher column density systems.
\mgtwo\ systems are clearly strongly related to the DLAs, but
likely occur at slightly larger impact parameters (Steidel 1993;
Ellison et al 2004b).  There is consequently no evidence to
suggest that the absorbers selected for this study should have
systematically different metallicities to the DLAs, although
extending this work with canonical DLA systems is desirable for
an entirely equal comparison.

The next important step in this work is to compare the emission and
absorption line abundances for a given galaxy.  Although we can
not make this comparison with our current data, we note that if QSO 
sightlines are truly a random probe of representative galaxy
metallicities at a given epoch, the apparent dichotomy between
emission and absorption abundances is intriguing (\textit{modulo}
the caveats above).
Is this dichotomy due to the different phases of the ISM probed
by emission and absorption line techniques?  Nebular abundances
within dwarf galaxies are usually highly uniform, implying that
local self-enrichment is not significant (e.g. Skillman 2003
and references therein).  We might therefore
expect emission and absorption abundances to be in good
agreement, at least on scales of a few kpc.  In the Milky Way, this is 
apparently the case:  oxygen
abundances in Orion's HII regions (e.g. Bautista \& Pradhan 1995)
are in good agreement with those measured from UV absorption lines in the
local interstellar medium (e.g. Moos et al.2002).  At high redshift,
Lyman break galaxies also exhibit nebular abundances commensurate
with those determined from absorption lines (e.g. Pettini et al.
2002).  There is some discussion concerning self-enrichment
in local dwarf galaxies (Aloisi et al. 2003; Lecavelier des Etangs 2004;
Lebouteiller et al. 2004), but since disagreement has arisen in
some cases for the same dataset, these results warrant further
investigation.

Ruling out convincing evidence for self-enrichment of HII regions
for the time being, we suggest that a possible explanation for a 
discrepancy between absorption and emission line abundances indicated 
by Figure 3 is a radial dependence on
metallicity.  Abundance gradients are commonplace in local spirals
(e.g. Vila-Costas \& Edmunds 1992; ZKH94) with magnitudes around
0.05--0.1 dex/kpc.  Studies of the extreme outer parts 
of disc galaxies find oxygen
abundances down to at least 1/10 to 1/15 of the solar value
(Ferguson et al. 1998; Kennicutt, Bresolin \& Garnett 2003), in good
agreement with DLA metallicities. 
Simulations of sightlines through galaxies with abundance gradients and
random inclinations can well reproduce the observed distribution
of DLA metallicities (Ferrini, Molla \& Diaz 1997; Mathlin et al. 2001), 
although 
diverse star formation histories and formation redshifts will 
also play a role.  The QSO impact parameters of the four galaxies
in our sample range from approximately 15 -- 50 \hkpc, so the
absorption abundances may be lower (by up to a few dex, depending on 
galaxy type, age and impact parameter) than in the global emission
line spectra which are representative of the central 5--10 kpc of a galaxy.
(Kobulnicky, Kennicutt \& Pizagno 1999). In \S3.1 (and
Figure 2) we showed that
DLA galaxies with absorption line metallicity determinations
lie below the LZ relation of $z<0.5$ galaxies.
If abundance gradients
existed in these four DLA galaxies with similar magnitudes as those
seen by Vila-Costas \& Edmunds (1992) and ZKH94, this would imply
an upward correction on the order of $\sim 1 - 1.5$ dex, bringing
these galaxies into good agreement with the field galaxy LZ
relation.  We will address the issue of abundance gradients
quantitatively in a forthcoming paper.

\section{Summary}

As stated in the introduction, the fundamental question which
we wish to address is whether QSO absorption and emission line
metallicity indicators provide the same abundance measurement
for a given galaxy.  As such, the work presented here is
merely a first step.  Our work shows that there is a discrepancy between 
absorption metallicities and the LZ relation for galaxies between $0.1<z<0.5$. 
For each Mg~II galaxy in our sample either (1) the metallicities 
lies on the $R_{23}$ 
upper branch, consistent with 
 the LZ relation, but significantly discrepant from the majority of DLA 
metallicities, or (2) the metallicity 
 lies on the lower branch, consistent with the DLA metallicities, but 
significantly different from 
 the LZ relation for emission-line galaxies (e.g. KZ99).   
If the metallicities lie on the upper branch, 
 a combination of abundance gradients and impact parameters is a 
possible resolution to 
 the discrepancy.  If the metallicities lie on the lower branch, 
this would suggest severe
 difference in the sample selection effects on metallicity between 
galaxies studied for the 
 LZ relation and absorption-selected galaxies.  
 
Ideally, one would like to compare emission and absorption-line 
abundances in the same DLA systems.
A direct comparison between emission and absorption line abundances
at low redshifts requires a combination of optical galaxy spectroscopy and 
UV QSO spectroscopy.   In order
to determine the absorption metallicity, it is preferable to target
a non-refractory $\alpha$ element so that a fair comparison can
be made with the nebular oxygen abundance.  Oxygen lines in the
UV are often problematic since they are either very strong and
consequently saturate even at low column densities (e.g. O~I
$\lambda$1302) or extremely weak (e.g. O~I $\lambda$1355).
The S~II $\lambda$1251, 1254, 1269  triplet presents a more 
viable option.  Bowen et al. (in preparation) have recently
determined the absorption metallicity for SBS~1543+593
via \textit{HST} spectroscopy and determine a sulphur abundance
that is in excellent agreement with the emission line
[O/H] measurement of Schulte-Ladbeck et al. (2004).  However,
this QSO has a very small impact parameter to the galaxy (0.5
\hkpc) and it remains to be seen how absorption and emission
line abundances compare for separations of 10 \hkpc\ or more.
Finally, we note that the concern of galaxy mis-identification
is inherent in this work, and the true absorbing galaxy may
be hidden by the glare of the QSO itself.  Using the optical
afterglows of gamma-ray bursts as a fading background
source (e.g. Jakobsson et al. 2004) with which to study intervening absorbers
is extremely promising in this regard.

\section*{Acknowledgments}

We are grateful to Marcin Sawicki for help with $k$-corrections and
Rolf Jansen for providing fits to the local galaxy $E(B-V)$ magnitude 
relation.  We have benefitted from discussions with, and
valuable advice from, Annette Ferguson, Margaret Geller, Chip 
Kobulnicky, Sandhya Rao, Max Pettini, Regina Schulte-Ladbeck
and Evan Skillman.

\end{document}